\documentclass{article}

\title{A general approach to physical realization of unambiguous quantum-state discrimination}
\author{Bing He, J\'{a}nos A. Bergou 
\thanks{Department of Physics and Astronomy, Hunter College of the City University of New York, 
 695 Park Avenue, New York, NY 10021, contact email: bhe98@earthlink.net}\\}

\begin{document}
\maketitle

\begin{abstract}
We present a general scheme to realize the POVMs for the unambiguous discrimination of quantum states. For any set of pure 
states it enables us to set up a feasible linear optical circuit to perform their optimal discrimination, if they are 
prepared as single-photon states. An example of unknown states discrimination is discussed as the illustration 
of the general scheme.

\vspace{0.2cm}
PACS number(s): 03.65.Bz, 03.67.-a, 03.65.Ta
\end{abstract}

Discrimination of the members of a set of $n$ quantum states $\{\rho_i\}$ ($i=1,2,\cdots,n$) is a standard task in quantum 
communication protocols. Typically these states are nonorthogonal pure states or even mixed states, so it is impossible to 
achieve one hundred percent correctness in the discrimination process. We can eliminate the chance of error but the price we will 
pay is that we must allow a chance of inconclusiveness. This senario is called unambiguous state discrimination (USD) 
\cite {ivanovic}, and it has found applications in quantum crytographic protocol  \cite {bennett} and quantum algorithm \cite {bergou}, etc. The physical methods 
proposed to do USD include linear quantum optics \cite {bergou2}, ion trap architecture \cite{roa} and nuclear magnetic 
resonance \cite {gopinath}. Some 
experiments on the USD of nonorthogonal photon states \cite {hutter} have been realized thus far. To perform the USD of nonorthogonal 
states, we need to use general positive operator-valued measures (POVMs) instead of orthogonal projectors, and their 
realization in the original signal Hilbert space of the measured system is usually impossible. Through Neumark's theorem \cite {neumark}, 
however, a POVM can be realized in the extended Hilbert spaces by performing unitary transformations and von Neumann 
projections together, and then the physical implementation of this POVM is feasible due to the fact that any discrete finite
dimensional unitary operator can be constructed in laboratory using optical devices \cite {reck}.
In finding the necessary unitary operators, we used to require the success probabilities, $p_i=Tr(\rho_i\Pi_k)\delta_{i,k}$, of the POVM elements $\Pi_k$ and the 
corresponding inconclusive probabilities, $q_i=1-p_i$ \cite {bergou3}. However, for the USD of a set of unknown states, 
we don't have the information available to obtain these probabilities and, therefore, are unable to find 
the required unitary transformations by the existing methods.

In this letter, we present a general scheme to realize {\it any} POVM for USD only with its 
elements $\Pi_k$ we have set up to unambiguously measure a particular set of pure or mixed inputs. 
The required unitary (orthogonal) transformations are found in an extended $2N$ dimensional Hilbert space if the dimension 
of the original signal Hilbert space is $N$. If the input states for discrimination are prepared as one-photon states, we can 
build a linear-optics circuit with at most $N^2$ lossless beam splitters to implement the POVM that unambiguously discriminates 
them. To reach the optimal performance of our set-up, we just need to choose the proper parameters 
(transmission and reflection coefficients) of the beam splitters such that the 
unitary (orthogonal) transformations in the extended space will achieve the same effect as
the optimal POVM does in the original space. It would also be equally possible to construct this device for the quantum 
states in the forms of any other type of systems, such as electron, neutron, atoms, etc. Moreover, this scheme is 
generalizable to the optical implementation of the Kraus operators of a POVM, and of any possible finite linear map given 
as a quantum operation, the general aspects of the physical realization for which has been discussed 
in \cite {dariano}.

The construction of the unitary transformations in the extended space by this general scheme is independent of the knowledge about the input 
states, so it can be applied to realize the USD of the unknown states, for which the inputs with certain symmetry are 
prepared with the copies of the unknown states as in \cite {bergou4, hayashi}. Take the inputs and the POVM in \cite {bergou4} for 
example, 
two unknown qubits $|\psi_{1}\rangle$ and $|\psi_{2}\rangle$, which are randomly distributed unit vectors on the Bloch sphere,
\begin{eqnarray}
|\psi_{i}\rangle=\cos(\theta_i/2)|0\rangle+\sin(\theta_i/2)
e^{i\varphi_i}|1\rangle\ ,
\end{eqnarray}
are prepared with their copies to produce the following quantum registers,
\begin{eqnarray}
|\Psi_{i}\rangle =  |\psi_{1}\rangle_{A}|\psi_{2}\rangle_{B}|\psi_{i}\rangle_{C}, 
\end{eqnarray}
where $i=1,2$,
with the probabilities of $\eta_1$ and $\eta_2$ repectively, and the POVM elements to unambiguously discriminate them 
are constructed as follows:
\begin{eqnarray}
\Pi_{1} = A_1^{\dagger}A_1&=& k_1 (P_{BC}^{as}\otimes I_A) \nonumber\\
\Pi_{2} = A_2^{\dagger}A_2&=& k_2 (P_{AC}^{as}\otimes I_B),
\end{eqnarray}
where $P_{BC}^{as}$ and $P_{AC}^{as}$ are the projectors into the anti-symmetric spaces of the respective digit locations.
The coefficients $k_1$ and $k_2$ can be properly tuned that this POVM will reach the optimum performance
in the USD of these inputs. Meanwhile, the probabilities of inconclusiveness are determined by the operator,
\begin{eqnarray}
\Pi_0=A^{\dagger}_{0}A_{0}=I-\Pi_{1}-\Pi_{2}. 
\end{eqnarray}
Since the inputs are unknown to us, the only quantity to indicate how well this POVM performs is the average succss 
probability. 
For the above example of 1 copy of $|\psi_{1}\rangle$ and $|\psi_{2}\rangle$ used as the reference in the registers, the maximum
average success probality  is $1/6$ if the
preparation probabilities $\eta_1$ and $\eta_2$ are equal \cite {bergou4}.

Generally, by a direct sum, we append an $M$ dimensional ancilla space ${\it A}$ to the $N$ dimensional original Hilbert space
${\it H}$ to realize a POVM through Neumark's theorem. With quantum optics techniques (see M. Mohseni {\it et al.} in \cite {hutter}), 
$n$ arbitrary initial input states $|\psi_{i}^{in}\rangle=
\Sigma_{j\in J}d_{i,j}a^{\dagger}_{j}|0\rangle$, where $J=\{1,2,\cdots,N+M\}$ and $d_{i,j}=0$ for $N+1\leq j\leq N+M$, are prepared 
as linear combinations of single-photon states through multi-rail representation \cite {milburn}, and are mapped by a unitary transformation in the extended space to the final states, 
the parts of which in the original Hilbert space ${\it H}$ are orthogonal and can be distinguished among themselves by 
counting the photons in different output ports. The output photons recorded in the ancilla ports then correspond to the 
inconclusive results. 

We start our direct sum realization of all such POVMs with the observation that the inconclusive operator $\Pi_0$ is 
positive and Hermitian in any of an orthonormal 
basis $\{|e_i\rangle\}$, where $i=1,2,\cdots,N$, in our $N$ dimensional Hilbert space $H$. Then we will find a unitary 
transformation $U$ that transforms the general orthonormal basis $\{|e_i\rangle\}$ to a unique orthonormal basis 
$\{|\alpha_i\rangle\}$ (up to some permutation), where $\Pi_0$ is diagonalized:
\begin{eqnarray}
U\Pi_0U^{\dagger}=\sum\limits_{i=1}^{N}c_i|\alpha_i\rangle\langle\alpha_i|.
\end{eqnarray}
Because $\Pi_0$ is a positive operator and $\langle\phi|\Pi_0|\phi\rangle\leq 1$ for any $|\phi\rangle$ in $H$, 
all its eigenvalues satisfy $0\leq c_i\leq 1$. From this fact we obtain the following
well-defined operators:
\begin{eqnarray}
A_0=A^{\dagger}_0=U^{\dagger}\left(\sum\limits_{i=1}^{N}\sqrt{c_i}|\alpha_i\rangle\langle\alpha_i|\right)U,
 \end{eqnarray}
 and
 \begin{eqnarray}
 (I-A_0^{\dagger}A_0)^{\frac{1}{2}}=U^{\dagger}\left(\sum\limits_{i=1}^{N}\sqrt{1-c_i}|\alpha_i\rangle\langle\alpha_i|\right)U,
 \end{eqnarray}
 if we represent them with the general orthonormal basis $\{|e_i\rangle\}$. With these operators we construct 
$\Sigma$ and other three unitary, or more exactly orthogonal transformation, operators in the extended $2N$ dimensional 
space as follows:
 \begin{eqnarray}
 \Sigma=\left(\begin{array}{cc}(I-A_0^{\dagger}A_0)^{\frac{1}{2}}& -A_0\\
 A_0 & (I-A_0^{\dagger}A_0)^{\frac{1}{2}}\\
\end{array}\right).
 \end{eqnarray}
The other three such operators are obtained by putting the minus sign in the upper right sub-matrix of $\Sigma$ to the other 
three sub-matrix blocks, respectively.
It is straightforward to prove $\Sigma^{\dagger}\Sigma=\Sigma\Sigma^{\dagger}=I$
with the operators defined in Eq. (6) and Eq. (7). In the most general 
situation when we have $n$ inputs to be unambiguously discriminated among themselves, $I-A_0^{\dagger}A_0$ in the square 
root equals 
$\Sigma_{i\in I}\Pi_i$, where $I=\{1,2,\cdots,n\}$.

We take $\Sigma$ to act on a set of states $\{\rho_i\}$ ($i=1,2,\cdots,n$), which are to be distinguished between each 
other in a USD process, in the extended $2N$ dimensional Hilbert space:
\begin{eqnarray}
 \Sigma\left(\begin{array}{cc}\rho_i & \\
  & o\\
\end{array}\right)\Sigma^{\dagger}
=\left(\begin{array}{cc}(I-A_0^{\dagger}A_0)^{\frac{1}{2}}\rho_i (I-A_0^{\dagger}A_0)^{\frac{1}{2}}& (I-A_0^{\dagger}A_0)^{\frac{1}{2}}\rho_i A_0^{\dagger}\\
 A_0\rho_i(I-A_0^{\dagger}A_0)^{\frac{1}{2}}& A_0\rho_iA_0^{\dagger}\\
\end{array}\right),
 \end{eqnarray}
 where the blank blocks and the $o$ sub-matrix represent the parts with all the entries $0$. The trace of the upper left diagonal block gives
 the success probability $p_i$ of unambiguously determining $\rho_i$ because $Tr(\rho_i\Pi_j)=p_i\delta_{i,j}$ for the
$\Pi_i$'s, and the trace of the lower right
 diagonal block gives the failure probability $q_i$ in the ancilla space $A$, and in the whole extended space
 $K=H\oplus A$ we have $p_i+q_i=Tr\rho_i=1$.
 
 If $\{\rho_i\}$ is a set of linearly independent pure states
$\{|\psi_i\rangle\}$, we will prove that the parts of their outputs after
 the action of $\Sigma$ are mutually orthogonal in the original signal Hilbert space $H$. Before the transformation, they 
are extended to the inputs $|\psi_i^{in}\rangle=(|\psi_i\rangle,{\bf 0})^{T}$,
 where $({\bf 0})$ represents a N-tuple of zero's, $(0,0,\cdots,0)$, in $A$. The output states are obtained as follows:
 \begin{eqnarray}
 |\psi_i^{out}\rangle&=&\left(\begin{array}{cc}(I-A_0^{\dagger}A_0)^{\frac{1}{2}}& -A_0\\
 A_0 & (I-A_0^{\dagger}A_0)^{\frac{1}{2}}\\
\end{array}\right)
\left(\begin{array}{c}|\psi_i\rangle\\
{\bf 0}\end{array}\right)\nonumber\\
&=&\left(\begin{array}{c}(I-A_0^{\dagger}A_0)^{\frac{1}{2}}|\psi_i\rangle\\
A_0|\psi_i\rangle \end{array}\right).
\end{eqnarray}
Then the inner product of the outputs for any pair of different $|\psi_i\rangle$ and $|\psi_j\rangle$ in $H$ 
is 
\begin{eqnarray}
\langle\psi_j|(I-A_0^{\dagger}A_0)^{\frac{1}{2}}(I-A_0^{\dagger}A_0)^{\frac{1}{2}}|\psi_i\rangle
=\langle\psi_j|\sum\limits_{k=1}^{n}A^{\dagger}_kA_k|\psi_i\rangle =0,
\end{eqnarray}
and in $A$ is
\begin{eqnarray}
\langle\psi_j|A^{\dagger}_0A_0 |\psi_i\rangle= \langle\psi_j|\psi_i\rangle.
\end{eqnarray}

For the mixed states we act $\Sigma$ on the products $\rho_i\rho_j$ ($i\neq j$) in the same way as in Eq. (9), and
find that the trace of the outputs in $H$ vanishes while in $A$ is the same as those of the inputs $\rho_i\rho_j$. 
Therefore, the unitary (orthogonal) transformation $\Sigma$ in the extended $2N$ dimensional Hilbert space realizes a scheme
to unambiguously discriminate any set of quantum states $\{\rho_i\}$.

Next we need to find the way of how to optically realize the unitary transformation $\Sigma$, given that the inputs are linear
combinations of single-photon states. Because the representation of $\Sigma$ may be complicated with the general
orthonormal basis, we apply the following unitary transformation
in $K$ to reduce it to be diagonalized in $4$ blocks:
\begin{eqnarray}
&&\left(\begin{array}{cc}U& \\
 & U\\
\end{array}\right)\left(\begin{array}{cc}(I-A_0^{\dagger}A_0)^{\frac{1}{2}}& -A_0\\
 A_0 & (I-A_0^{\dagger}A_0)^{\frac{1}{2}}\\
\end{array}\right)\left(\begin{array}{cc}U^{\dagger}& \\
 & U^{\dagger}\\
\end{array}\right)\nonumber\\
&=&\left(\begin{array}{cc}\sum\limits_{i=1}^{N}\sqrt{1-c_i}|\alpha_i\rangle\langle\alpha_i|& -\sum\limits_{i=1}^{N}\sqrt{c_i}|\alpha_i\rangle\langle\alpha_i|\\
 \sum\limits_{i=1}^{N}\sqrt{c_i}|\alpha_i\rangle\langle\alpha_i|& \sum\limits_{i=1}^{N}\sqrt{1-c_i}|\alpha_i\rangle\langle\alpha_i|\\
\end{array}\right).
\end{eqnarray} 
We use $T_{p,q}$ to define an identity matrix with the entries $I_{pp}$, $I_{pq}$, $I_{qp}$ and $I_{qq}$ replaced by the corresponding
$O(2)$ matrix elements. If we multiply the matrix obtained in Eq. (13) by $T_{p,q}$, only the entries at $(p,p)$, $(p,q)$, $(q,p)$
and $(q,q)$ positions will be changed with all other entries intact. Taking out the entries on the $i$th and $(i+N)$th
rows and columns for example, and multiplying them on the right by the corresponding $O(2)$ sub-matrix in $T_{i+N,i}$, we have
\begin{eqnarray}
&&\left(\begin{array}{cc}\sqrt{1-c_i}&-\sqrt{c_i} \\
 \sqrt{c_i}& \sqrt{1-c_i}\\
\end{array}\right)\left(\begin{array}{cc}\cos\theta_i& -\sin\theta_i\\
\sin\theta_i & \cos\theta_i\\
\end{array}\right)\nonumber\\
&=&\left(\begin{array}{cc}\cos\theta_i\sqrt{1-c_i}-\sin\theta_i\sqrt{c_i}& -\sin\theta_i\sqrt{1-c_i}-\cos\theta_i\sqrt{c_i}\\
\sin\theta_i \sqrt{1-c_i}+\cos\theta_i\sqrt{c_i}&\cos\theta_i\sqrt{1-c_i}-\sin\theta_i\sqrt{c_i} \\
\end{array}\right).~~
\end{eqnarray} 
If we choose $\tan\theta_i=-\sqrt{\frac{c_i}{1-c_i}}$ to let the off-diagonal elements vanish, the diagonal elements will
be $1$ and an identity matrix will be obtained. This two dimensional rotational transformation matrix, with its diagonal and
 off-diagonal elements given as the transmission and reflection coefficient respectively, can be implemented by a lossless 
beam splitter, if the inputs are single-photon states. Similarly, if we right-multiply the whole matrix successively by 
$T_{2N,N}$,$T_{2N-1,N-1}$, $\cdots$, $T_{N+1,1}$, we will obtain
\begin{eqnarray}
\left(\begin{array}{cc}\sum\limits_{i=1}^{N}\sqrt{1-c_i}|\alpha_i\rangle\langle\alpha_i|& -\sum\limits_{i=1}^{N}\sqrt{c_i}|\alpha_i\rangle\langle\alpha_i|\\
 \sum\limits_{i=1}^{N}\sqrt{c_i}|\alpha_i\rangle\langle\alpha_i|& \sum\limits_{i=1}^{N}\sqrt{1-c_i}|\alpha_i\rangle\langle\alpha_i|\\
\end{array}\right)T_{2N,N}T_{2N-1,N-1}\cdots T_{N+1,1}=I,
\end{eqnarray} 
an identity matrix in the extended $2N$ dimensional space $K$. Therefore $\Sigma$ represented by the basis 
$\{|\alpha_i\rangle\}$ is given as
\begin{eqnarray}
\Sigma=T_{N+1,1}^{-1}T_{N+2,2}^{-1}\cdots T_{2N,N}^{-1},
\end{eqnarray} 
and can be exactly implemented by $N$ lossless beam splitters \cite {reck}. Following the notation of P. Kok {\it et al.} 
in \cite {reck}, we choose
the relative phase shift $\varphi=\frac{\pi}{2}$ for these beam splitters. 

In most realistic applications, we also need a post-process unitary transformation $V$ after the action of $\Sigma$
so that the output states, which have already been orthogonal, will be mapped to different final output ports for measurement. 
Together with
the former procedure, the whole unitary transformation $U_T$ (in the extended space $K$) required to realize the USD of a set of inputs 
is constructed as follows:
\begin{eqnarray}
U_T=\left(\begin{array}{cc}V& \\
 & V\\
\end{array}\right)\left(\begin{array}{cc}\sum\limits_{i=1}^{N}\sqrt{1-c_i}|\alpha_i\rangle\langle\alpha_i|& -\sum\limits_{i=1}^{N}\sqrt{c_i}|\alpha_i\rangle\langle\alpha_i|\\
 \sum\limits_{i=1}^{N}\sqrt{c_i}|\alpha_i\rangle\langle\alpha_i|& \sum\limits_{i=1}^{N}\sqrt{1-c_i}|\alpha_i\rangle\langle\alpha_i|\\
\end{array}\right)\left(\begin{array}{cc}U& \\
 & U\\
\end{array}\right),
\end{eqnarray}
which consists of the pre-process transformation $U$, $\Sigma$ in the form of $4$ diagonalized sub-matrices and the post-process
transformation $V$. We don't actually need $U$ and $V$ in the ancilla space $A$, so the total number of the beam splitters
required to perform the USD of the inputs has an upper bound $2\times (N(N-1)/2)+N=N^2$, a value determined by the dimension
of the Hilbert space of the measured system. 

As an application of this general scheme, we show how to realize the USD of a pair of unknown qubits discussed in \cite {bergou4}. 
We assume that the preparation probabilities $\eta_1$ and $\eta_2$ of the inputs are equal for simplicity (the unequal 
$\eta_i$'s situation is realized in the same way but the matrix elements involved would look more complicated) 
and give the optical realization procedure to achieve their optimal USD. 

The quantum registers in Eq. (2) are expanded as $|\Psi_i\rangle=\Sigma_x d_{i,x}|x \rangle$, where $\{|x\rangle\}$ is
$\{|000\rangle,|001\rangle,\cdots,|111\rangle\}$ and can be realized by $8$ modes of single-photon state $|k\rangle=a_k^{\dagger}|0\rangle
$ ($k=1,2,\cdots,8$) respectively. With these basis vectors we perform the following transformations:
\begin{eqnarray}
|\eta_{i1}\rangle&=&\sqrt{\frac{1}{2}}|i10\rangle+\sqrt{\frac{1}{2}}|i01\rangle \nonumber\\
|\chi_{i1}\rangle&=&\sqrt{\frac{1}{2}}|i10\rangle-\sqrt{\frac{1}{2}}|i01\rangle,
\end{eqnarray}
where $i=0,1$,
to map them to a set of another orthonormal basis 
\begin{eqnarray}
\{|g_i\rangle\}=\{|000\rangle,|100\rangle,|\eta_{01}\rangle,|\chi_{01}\rangle,
|011\rangle,|\eta_{11}\rangle,|\chi_{11}\rangle,|111\rangle\}.
\end{eqnarray}
These orthogonal transformations can be implemented by beam splitters in the input states preparation period, and
the advantage of doing this is that the multiplicity of the eigenvalues of $\Pi_0$ will be conveniently demonstrated if we represent
it with this basis. Then the quantum registers for the USD process are represented with this basis as follows:
\begin{eqnarray}
|\Psi_i\rangle &=&|\Psi_i^{(0)}\rangle +|\Psi_i^{(1)}\rangle +|\Psi_i^{(2)}\rangle+|\Psi_i^{(3)}\rangle,
\end{eqnarray}
where
\begin{eqnarray}
|\Psi_1^{(1)}\rangle &=&
\left( \begin{array}{c} \cos\frac{\theta_1}{2}\sin\frac{\theta_1}{2}e^{i\varphi_1}\cos\frac{\theta_2}{2}\\
\frac{1}{\sqrt{2}}\cos\frac{\theta_1}{2}\sin\frac{\theta_1}{2}e^{i\varphi_1}\cos\frac{\theta_2}{2}+\frac{1}{\sqrt{2}}
\cos^2\frac{\theta_1}{2}\sin\frac{\theta_2}{2}e^{i\varphi_2}\\
-\frac{1}{\sqrt{2}}\cos\frac{\theta_1}{2}\sin\frac{\theta_1}{2}e^{i\varphi_1}\cos\frac{\theta_2}{2}+\frac{1}{\sqrt{2}}
\cos^2\frac{\theta_1}{2}\sin\frac{\theta_2}{2}e^{i\varphi_2}
\end{array}\right),\nonumber\\
|\Psi_2^{(1)}\rangle &=&\left( \begin{array}{c} \sin\frac{\theta_1}{2}e^{i\varphi_1}\cos^2\frac{\theta_2}{2}\\
\sqrt{2}\cos\frac{\theta_1}{2}\cos\frac{\theta_2}{2}\sin\frac{\theta_2}{2}e^{i\varphi_2}\\
0
\end{array}\right),
\end{eqnarray}
represented by $\{|100\rangle,|\eta_{01}\rangle,|\chi_{01}\rangle\}$, and $|\Psi_1^{(2)}\rangle $ and
$|\Psi_2^{(2)}\rangle $, which are spanned by the set of basis vectors $\{|011\rangle,
|\eta_{11}\rangle,|\chi_{11}\rangle\}$, just 
carry the above $|\Psi_1^{(1)}\rangle$,$|\Psi_2^{(1)}\rangle$ components but with
one $\cos\frac{\theta_i}{2}$ in the terms replaced by $\sin\frac{\theta_i}{2}e^{i\varphi_i}$ because
there is one more digit $1$ in the basis vectors. The $|\Psi_1^{(0)}\rangle$ and $|\Psi_1^{(3)}\rangle$
(resp. $|\Psi_2^{(0)}\rangle$ 
 and $|\Psi_2^{(3)}\rangle$) in the joint space of $|\Psi_1\rangle$ and $|\Psi_2\rangle$
are given as
\begin{eqnarray}
|\Psi_1^{(0)}\rangle &=& \cos^2\frac{\theta_1}{2}\cos\frac{\theta_2}{2}|000\rangle \nonumber\\
|\Psi_1^{(3)}\rangle &=& \sin^2\frac{\theta_1}{2}e^{i2\varphi_1}\sin\frac{\theta_2}{2}e^{i\varphi_2}|111\rangle, 
\end{eqnarray}
and $|\Psi_2^{(0)}\rangle$, $|\Psi_2^{(3)}\rangle$ are obtained by interchanging $\theta_1$, $\phi_1$ with 
$\theta_2$, $\phi_2$ for all the factors in Eq. (22). We should bear in mind that we know nothing about $\theta_i$'s and $\varphi_i$'s because
$|\psi_1\rangle$ and $|\psi_2\rangle$ are randomly distributed on the Bloch sphere.

With the basis $\{|g_i\rangle\}$ the inconclusive operator $\Pi_0$ for the optimal USD measurement is given as the following
direct sum:
\begin{eqnarray}
\Pi_0=\Gamma_0\oplus \Gamma_1\oplus \Delta_1\oplus \Delta_0,
\end{eqnarray}
where $\Gamma_0$ and $\Delta_0$ are $1\times 1$ matrices $1$ acting on the spaces spanned by $|000\rangle$ and $|111\rangle$, 
respectively, 
and 
\begin{eqnarray}
\Gamma_1=\left(\begin{array}{ccc}
\frac{2}{3}&\frac{\sqrt{2}}{6}& -\frac{\sqrt{2}}{6}\\
\frac{\sqrt{2}}{6}&\frac{5}{6}&\frac{1}{6}\\
-\frac{\sqrt{2}}{6}&\frac{1}{6}&\frac{1}{6}\\
\end{array}\right)
\end{eqnarray}\
acting on the space spanned by $\{|100\rangle,|\eta_{01}\rangle,|\chi_{01}\rangle\}$, and $\Delta_1$ acting on the space spanned
by $\{|011\rangle,|\eta_{11}\rangle,|\chi_{11}\rangle\}$ looks similar to $J_1$ with the signs of some off-diagonal elements
changed. There is a rotation, 
\begin{eqnarray}
U^{(1)}&=&\left(\begin{array}{ccc}
\frac{\sqrt{6}}{6}&-\frac{\sqrt{3}}{6}& \frac{\sqrt{3}}{2}\\
-\frac{\sqrt{2}}{2}&\frac{1}{2}&\frac{1}{2}\\
\frac{\sqrt{3}}{3}&\frac{\sqrt{6}}{3}&0\\
\end{array}\right),
\end{eqnarray}
which can be decomposed into three $2$ dimensional rotations and is thus realized by three beam splitters together, to get 
$\Gamma_1$ diagonalized:
\begin{eqnarray}
\Gamma_1'=U^{(1)}\Gamma_1U^{(1)T}=\left(\begin{array}{ccc}
0&0& 0\\
0&\frac{2}{3}&0\\
0&0&1\\
\end{array}\right),
\end{eqnarray}
and $\Delta_1$ is diagonalized to the same matrix or has the same eigenvalues as $\Gamma_1$. Now we have
the diagonalized operators,
\begin{eqnarray}
(I-\Gamma_1')^{\frac{1}{2}}=(I-\Delta_1')^{\frac{1}{2}}=\left(\begin{array}{ccc}
1&0& 0\\
0&\sqrt{\frac{1}{3}}&0\\
0&0&0\\
\end{array}\right),
\end{eqnarray}
acting on the respective sub-spaces of $H$, and the unitary (orthogonal) operator
\begin{eqnarray}
\Sigma=\left(\begin{array}{cccccccc}
0 & & & & -1 & & & \\
 &(I-\Gamma_1')^{\frac{1}{2}}& & &  &-(\Gamma_1')^{\frac{1}{2}} & & \\
& &(I-\Delta_1')^{\frac{1}{2}}& & & & -(\Delta_1')^{\frac{1}{2}}& \\
& & & 0 & & & & -1\\ 
1 & & & & 0 & & &\\
 &(\Gamma_1')^{\frac{1}{2}}& & &  &(I-\Gamma_1')^{\frac{1}{2}} & & \\\\
& &(\Delta_1')^{\frac{1}{2}}& & & & (I-\Delta_1')^{\frac{1}{2}}& \\
& & & 1 & & & & 0\\ 
\end{array}\right),
\end{eqnarray}
acting on the extended space $K$.

We therefore just need to look at the USD of $|\Psi_1^{(1)}\rangle$ and $|\Psi_2^{(1)}\rangle$ for the whole problem.
The output parts of these states in the original Hilbert space after $U$ and $\Sigma$ are
\begin{eqnarray}
|\Psi_1^{(1)\prime}\rangle &=&(I-\Gamma_1')^{\frac{1}{2}}U^{(1)}|\Psi_1^{(1)}\rangle =
\left( \begin{array}{c} -\frac{1}{\sqrt{6}}\cos\frac{\theta_1}{2}\sin\frac{\theta_1}{2}e^{i\varphi_1}\cos\frac{\theta_2}{2}
+\frac{1}{\sqrt{6}}\cos^2\frac{\theta_1}{2}\sin\frac{\theta_2}{2}e^{i\varphi_2}\\
-\frac{1}{\sqrt{6}}\cos\frac{\theta_1}{2}\sin\frac{\theta_1}{2}e^{i\varphi_1}\cos\frac{\theta_2}{2}
+\frac{1}{\sqrt{6}}\cos^2\frac{\theta_1}{2}\sin\frac{\theta_2}{2}e^{i\varphi_2}\\
0
\end{array}\right),\nonumber\\
|\Psi_2^{(1)\prime}\rangle &=&(I-\Gamma_1')^{\frac{1}{2}}U^{(1)}|\Psi_2^{(1)}\rangle=\left( \begin{array}{c} 
\frac{1}{\sqrt{6}}\sin\frac{\theta_1}{2}e^{i\varphi_1}\cos^2\frac{\theta_2}{2}-\frac{1}{\sqrt{6}}\cos\frac{\theta_1}{2}
\cos\frac{\theta_2}{2}\sin\frac{\theta_2}{2}e^{i\varphi_2}\\
-\frac{1}{\sqrt{6}}\sin\frac{\theta_1}{2}e^{i\varphi_1}\cos^2\frac{\theta_2}{2}+\frac{1}{\sqrt{6}}\cos\frac{\theta_1}{2}
\cos\frac{\theta_2}{2}\sin\frac{\theta_2}{2}e^{i\varphi_2}\\
0
\end{array}\right),~~~~~~
\end{eqnarray}
respectively. At this step they have been orthogonal to each other, and then we add a Hadamard gate, 
\begin{eqnarray}
V^{(1)}=\left(\begin{array}{ccc}
\frac{1}{\sqrt{2}}&\frac{1}{\sqrt{2}}& 0\\
\frac{1}{\sqrt{2}}&-\frac{1}{\sqrt{2}}&0\\
0&0&1\\
\end{array}\right),
\end{eqnarray}
as the post-process to rotate these two states such that they will have non-zero components
in different output ports. Thus the final states will be unambiguously distinguished between each other once the photons
are recorded in different output ports. 

The process on $|\Psi_1^{(2)}\rangle$ and $|\Psi_2^{(2)}\rangle$ is similar
too. $|\Psi_1^{(0)}\rangle$, $|\Psi_2^{(0)}\rangle$ and $|\Psi_1^{(3)}\rangle$, $|\Psi_2^{(3)}\rangle$ will be zero vectors
after we map them with the corresponding transformations to the final states. Together with $2$ zero components produced
in other sub-spaces like those in Eq. (29),
there is a $4$ dimensional joint space of $|\Psi_{1}\rangle$ and $|\Psi_{2}\rangle$, the vector components in which
only contribute to the inconclusive result. The average of the inner products of the outpout vectors in Eq. (29) is $1/12$.
Putting it together with the corresponding contribution from $|\Psi_1^{(2)}\rangle$ or $|\Psi_2^{(2)}\rangle$, we have
the total average success probability $1/6$ for the optimal USD of $|\Psi_1\rangle$ and $|\Psi_2\rangle$  \cite {bergou4}.

All the transformations in the process only need $14$ beam splitters (including some totally
 reflecting mirrors) to implement. Moreover, in the whole
process, we don't need to acquire any information about $\theta_i$'s and $\varphi_i$'s of the randomly distributed qubits.
By the general scheme we propose in this paper, we realize the USD of the unknown states prepared as in Eq. (2).

B. He would like to thank Dr. Z. Wang for helpful discussions.

\bibliographystyle{unsrt}

\end{document}